\documentclass[aps,prd,onecolumn,groupedaddress,showpacs,nofootinbib,amssymb,preprintnumbers]{revtex4}
\usepackage[T1]{fontenc}
\usepackage[latin1]{inputenc}
\usepackage{graphicx}
\usepackage[english]{babel}
\usepackage{amsmath}
\usepackage{amssymb}
\usepackage{amsfonts}

\begin{document}

\newcommand{\be}{\begin{equation}}
\newcommand{\ee}{\end{equation}}
\newcommand{\bea}{\begin{eqnarray}}
\newcommand{\eea}{\end{eqnarray}}
\newcommand{\beaa}{\begin{eqnarray*}}
\newcommand{\eeaa}{\end{eqnarray*}}
\newcommand{\Lhat}{\widehat{\mathcal{L}}}
\newcommand{\nn}{\nonumber \\}
\newcommand{\e}{\mathrm{e}}
\newcommand{\tr}{\mathrm{tr}\,}

\tolerance=5000

\preprint{AEI-2011-051}

\title{Cyclic, ekpyrotic and little rip universe in modified gravity}

\author{
Shin'ichi Nojiri$^1$\footnote{nojiri@phys.nagoya-u.ac.jp},
Sergei D. Odintsov$^{2,3}$\footnote{odintsov@ieec.uab.es}$^{,}$\footnote{Also
at Tomsk State Pedagogical University.},
and D.~S\'{a}ez-G\'{o}mez$^3$\footnote{saez@ieec.uab.es}}
\affiliation{$^1$Department of Physics, Nagoya University, Nagoya 464-8602,
Japan \\
and Kobayashi-Maskawa Institute for the Origin of Particles and the Universe,
Nagoya University, Nagoya 464-8602, Japan}
\affiliation{$^2$Instituci\`{o} Catalana de Recerca i Estudis Avan\c{c}ats
(ICREA), Barcelona, Spain}
\affiliation{$^3$Institut de Ci\`encies de l'Espai
ICE (CSIC-IEEC), Campus UAB Facultat de Ci\`encies, Torre C5-Parell-2a
pl, E-08193 Bellaterra (Barcelona) Spain}

\begin{abstract}

We propose the reconstruction of $F(R)$ gravity in such a way that
corresponding theory admits cyclic and ekpyrotic universe solutions.
The number of explicit examples of such $F(R)$ model is found.
The comparison with the reconstructed scalar-tensor theory is made.
We also present $F(R)$ gravity which provides the little rip evolution and
gives the realistic gravitational alternative for $\Lambda$CDM cosmology.
The time for little rip dissolution of bound structures in such theory is
estimated.
We demonstrate that transformed little rip $F(R)$ solution becomes
qualitatively
different cosmological solution with Big Bang type singularity in Einstein
frame.

\end{abstract}

\pacs{95.36.+x, 98.80.Cq}

\maketitle

\section{Introduction}

It is widely accepted nowadays that universe evolution
has passed through two (super)accelerating epochs: early-time inflation and
late-time cosmic acceleration. Despite
the close similarity between these two accelerating epochs there is a
number of differences. For instance, inflationary era is
characterized by large value of curvature while dark era occurs
at small curvature. Moreover, the effective equation of state parameter $w$ is
qualitatively different for inflation or late-time
cosmic acceleration. Inflation occurs right after
Big Bang/Big Crunch singularity while number of
quintessence/phantom dark energy models end up at finite time
future singularity. This suggests the very natural conjecture:
Big Bang should be identified with finite time singularity.
In other words, the universe realizes cyclic evolution
which may be produced by multi-fluid \cite{paul} or
by the theory with oscillating equation of
state parameter \cite{OscillatingUn}. It is natural to construct
the cyclic universe within the theory which describes the inflation and dark
energy in unified way.

Modified gravity is realistic alternative for unified
description of inflation with dark energy (for review of such
unification in modified gravity, see \cite{review}). Moreover,
its weak-field limit may lead to newtonian regime (for review,
see \cite{faraoni}). Hence, modified gravity represents very natural candidate
where cyclic cosmology may be realized.
The purpose of this work is precisely this one: reconstruction of modified
$F(R)$ gravity which leads to cyclic evolution. In the next section the
reconstruction technique for scalar-tensor theory and $F(R)$ gravity is
developed. We find the explicit examples of scalar-tensor theory and $F(R)$
gravity which have the solutions corresponding to cyclic universe.
In third section $F(R)$ action is presented as General Relativity plus ideal
fluid. The reconstruction is developed for such formulation of modified
gravity. Again, the example of $F(R)$ gravity which admits cyclic universe
solution is found. In fourth section using above methods we show that ekpyrotic
scenario may be also realized in frames of modified gravity.
In fifth section we reconstruct $F(R)$ gravity which induces little rip
cosmology. Such non-singular dark energy model was recently
proposed in Ref.~\cite{LittleRip} as an
alternative to $\Lambda$CDM model. Furthermore, we demonstrate that
reconstruction of $F(R)$ theory which leads to non-singular phantom
evolution is possible.
Some summary is given in the last section.

\section{Conformally cyclic universe}

Let us show how to construct the cyclic universe solution in modified gravity.
The starting point is the spatially flat FRW metric
\be
\label{I}
ds^2 = - dt^2 + a(t)^2 \sum_{i=1,2,3} \left( dx^i\right)^2 \, .
\ee
where $a(t)$ is the monotonically expanding function of the cosmological time
$t$.
Let $a(t)$ has the following form
\be
\label{II}
\ln a(t) = H_0 t + h(t)\, .
\ee
Here $H_0$ is a positive constant and $h(t)$ is a function with period $T$:
\be
\label{III}
h(t+T) = h(t)\, .
\ee
Then
\bea
\label{IV}
ds^2 &=& - dt^2 + a(t+T)^2 \sum_{i=1,2,3} \left( dx^i\right)^2 \nn
&=& -dt^2 + a(t)^2 \e^{2H_0 T} \sum_{i=1,2,3} \left( dx^i\right)^2 \nn
&=& -dt^2 + a(t)^2 \sum_{i=1,2,3} \left( d{\tilde x}^i\right)^2 \, .
\eea
Here $\tilde x^i \equiv \e^{H_0 T} x^i$.
Eqs.~(\ref{I}) and (\ref{IV}) show that the metric at $t$ is
physically identical with that at $t+T$.
In this sense, the universe is cyclic although the universe is
monotonically expanding. Hence, it called
as conformally cyclic universe.
A simplest example is
\be
\label{V}
h(t) = H_1 \cos \left( \frac{2\pi t}{T} \right)\, .
\ee
Here $H_1$ is a positive constant less than $H_0$, which guarantees that the
universe is monotonically expanding.

We now consider the models which reproduce the scale factor $a(t)$ given by
(\ref{II}) with (\ref{V}).
First, let us investigate the scalar-tensor model, whose action is given by
\be
\label{ma7}
S=\int d^4 x \sqrt{-g}\left\{
\frac{1}{2\kappa^2}R - \frac{1}{2}\omega(\phi)\partial_\mu \phi
\partial^\mu\phi - V(\phi) + L_\mathrm{matter} \right\}\, .
\ee
Here, $\omega(\phi)$ and $V(\phi)$ are functions of the scalar $\phi$.
For the model where $\omega(\phi)$ and $V(\phi)$ are given by a single function
$f(\phi)$, as follows,
\be
\label{ma10}
\omega(\phi)=- \frac{2}{\kappa^2}f''(\phi)\, ,\quad
V(\phi)=\frac{1}{\kappa^2}\left(3f'(\phi)^2 + f''(\phi)\right)\, ,
\ee
the exact solution of the FRW equations has the following form:
\be
\label{ma11}
\phi=t\, ,\quad H=f'(t)\, .
\ee
In case of (\ref{II}) with (\ref{V}), one finds
\be
\label{VI}
f(t) = \ln a(t)\, , \quad f'(t) = H_0 - \frac{2\pi H_1}{T} \sin \left(
\frac{2\pi t}{T}\right)\, .
\ee
Therefore if we consider the model
\bea
\label{VII}
\omega(\phi) &=& \frac{2 H_1}{\kappa^2} \left(\frac{2\pi}{T}\right)^2
\cos \left( \frac{2\pi \phi}{T} \right)\, ,\nn
V(\phi) &=& \frac{1}{\kappa^2}\left\{3 \left( H_0 - \frac{2\pi H_1}{T}
\sin \left( \frac{2\pi \phi}{T}\right) \right)^2
 - \left(\frac{2\pi}{T}\right)^2 \cos \left( \frac{2\pi \phi}{T} \right)
\right\}\, ,
\eea
the conformal cyclic universe (\ref{II}) with (\ref{V}) can be reproduced.

We now consider $F(R)$ theory
\be
\label{JGRG7}
S_{F(R)}= \int d^4 x \sqrt{-g} \left( \frac{F(R)}{2\kappa^2}
+ \mathcal{L}_\mathrm{matter} \right)\, .
\ee
Here $F(R)$ is an appropriate function of the scalar curvature $R$.
The action (\ref{JGRG7}) is equivalently rewritten as
\be
\label{PQR1}
S=\int d^4 x \sqrt{-g} \left\{\frac{1}{2\kappa^2} \left( P(\phi) R
+ Q(\phi) \right) + \mathcal{L}_\mathrm{matter}\right\}\, .
\ee
Here, $P$ and $Q$ are proper functions of the auxiliary scalar $\phi$.
By the variation over $\phi$, it follows that
$0=P'(\phi)R + Q'(\phi)$,
which may be solved with respect to $\phi$ as $\phi=\phi(R)$.
By substituting the obtained expression of $\phi(R)$ into (\ref{PQR1}),
one arrives again at the $F(R)$ gravity action (for reconstruction of modified
gravities,
see Refs.~\cite{review,Reconstruction,Nojiri:2009kx}):
\be
\label{PQR4}
S=\int d^4 x \sqrt{-g} \left\{\frac{F(R)}{2\kappa^2}
+ \mathcal{L}_\mathrm{matter}\right\}\, , \quad
F(R)\equiv P(\phi(R)) R + Q(\phi(R))\, .
\ee
If the scale factor $a$ is
given by a proper function $g(t)$ as $a=a_0\e^{g(t)}$ with a constant $a_0$,
by solving the following second-rank differential equation,
\be
\label{PQR11}
0 = 2 \frac{d^2 P(\phi)}{d\phi^2} - 2 g'(\phi) \frac{dP(\phi))}{d\phi}
+ 4g''(\phi) P(\phi) + \sum_i \left(1 + w_i\right) \rho_{i0} a_0^{-3(1+w_i)}
\e^{-3(1+w_i)g(\phi)} \, ,
\ee
we can find the form of $P(\phi)$ which gives a solution $a=a_0\e^{g(t)}$.
One also finds the form of $Q(\phi)$ as follows:
\be
\label{eq:2.12}
Q(\phi)=-6 \left(g'(\phi)\right)^2 P(\phi) - 6g'(\phi)
\frac{dP(\phi)}{d\phi} \nn
+ \sum_i \rho_{i0} a_0^{-3(1+w_i)} \e^{-3(1+w_i)g(\phi)} \, .
\ee
In case of (\ref{II}) with (\ref{V}), when we neglect the contribution from
matter, Eq.~(\ref{PQR11}) has the following form:
\be
\label{VIII}
0 = \frac{d^2 P(\phi)}{d\phi^2} - \left( H_0
 - \frac{2\pi H_1}{T} \sin\left( \frac{2\pi\phi}{T} \right) \right)
\frac{dP(\phi))}{d\phi}
 - \frac{2\left(2\pi\right)^2}{T^2} \cos\left( \frac{2\pi\phi}{T} \right)
P(\phi) \, .
\ee
Since one cannot solve (\ref{VIII}) explicitly, we assume $P(\phi)$ can be
expanded by the Fourier series:
\be
\label{IX}
P(\phi) = \sum_{n=0}^\infty p_n \cos \left( \frac{2\pi \phi}{T} \right)
+ \sum_{n=1}^\infty q_n \sin \left( \frac{2\pi \phi}{T} \right) \, .
\ee
Then one gets $q_1 = p_1 =0$ and for $n\geq 2$,
\bea
\label{X}
p_n &=& - \frac{2\left(n-1\right)^2}{H_1 \left(n+2\right)} p_{n-1}
 - \frac{4\pi H_0 \left(n-1\right)}{T \left(n+2\right)} q_{n-1}
+ \frac{n-4}{n+2} p_{n-2} \, ,\nn
q_n &=& \frac{2\left(n-1\right)^2}{H_1 \left(n+2\right)} q_{n-1}
 - \frac{4\pi H_0 \left(n-1\right)}{T \left(n+2\right)} p_{n-1}
 - \frac{n-4}{n+2} q_{n-2} \, .
\eea
In (\ref{X}), we put $q_0 = 0$.
Eq.~(\ref{X}) shows the existence of the solution of (\ref{VIII}) and the
corresponding $F(R)$ gravity model.

When matter can be neglected, Eq.~(\ref{PQR11}) can be rewritten as
\be
\label{XI}
\frac{d}{d\phi} \left( g'(\phi) P(\phi)^{-1/2} \right)
= - \frac{1}{2} P(\phi)^{-3/2} \frac{d^2 P(\phi)}{d\phi^2}\, ,
\ee
which gives
\be
\label{XII}
g'(\phi) = - \frac{1}{2} P(\phi)^{1/2} \int d\phi P(\phi)^{-3/2} \frac{d^2
P(\phi)}{d\phi^2}
= - \frac{1}{2P(\phi)} \frac{d P(\phi)}{d\phi} - \frac{3}{4}P(\phi)^{1/2} \int
d\phi
P(\phi)^{-5/2} \left( \frac{d P(\phi)}{d\phi} \right)^2 \, .
\ee
In the second equality, we have used the partial integration.
Furthermore by writing $P(\phi)$ as
\be
\label{XIII}
P(\phi) = U(\phi)^{-4} \, ,
\ee
(\ref{XII}) is rewritten as follows:
\be
\label{XIV}
g'(\phi) = \frac{2}{U(\phi)}\frac{d U(\phi)}{d\phi} - \frac{12}{U(\phi)^2}
\int d\phi \left( \frac{d U(\phi)}{d\phi} \right)^2\, .
\ee
As an example, we may consider
\be
\label{XX}
U(\phi) = U_1 + U_2 \cos \omega \phi\, .
\ee
Here $U_1$, $U_2$, and $\omega$ are constants and it is assumed $U_1>U_2>0$,
which show $U(\phi)$ does not vanish.
Hence
\be
\label{XXI}
g'(\phi) = - \frac{ 2 U_2 \omega \sin \omega \phi}{U_1 + U_2 \cos \omega \phi}
 - \frac{3 U_2^2 \omega^2 \left( 2 \phi - \frac{1}{\omega} \sin \left(2\omega
\phi\right) + C \right)}{\left( U_1 + U_2 \cos \omega \phi \right)^2 }\, .
\ee
Here $C$ is a constant of the integration.
Note that the obtained $g'(\phi)$ is not always positive. We also note that the
$g'(\phi)$ is not periodic
due to the term proportional to $\phi$. This kind of term always appears for
the periodic $U(\phi)$ since the
integrand of the last term in (\ref{XIV}) is positive definite and therefore
the integration cannot be periodic.

Let us now consider an example where the function $P(\phi)$ is given by,
\be
\label{dd2}
P(\phi) = U(\phi)^{-4}=P_0 (\cos \omega \phi)^4 \, ,
\ee
where $P_0$ and $\omega$ are constants. Then, by the equation (\ref{XI}), the
solution is,
\be
g'(\phi)=g_0(\cos \omega \phi)^2+2\omega (\sin 2 \omega \phi-\tan \omega
\phi)\, ,
\label{dd3}
\ee
where $g_0$ is an integration constant. Note that here the solution contains
divergences due to the term
of the tangent in (\ref{dd2}). These divergences correspond to points where the
scale factor becomes
null $a(t_0)=0$, which can be identified with a Big Bang/Crunch singularity.
This class of singularities is very common in cyclic Universes,
where the ekpyrotic scenario is reproduced. In the same way other cyclic
universes may be used for explicit reconstruction of modified gravity.

\section{ Reconstructed $F(R)$ as an effective perfect fluid}

Let us start with the modified gravity FRW equations written in the following
form:
\bea
3H^2 &=& \frac{1}{F'(R)}\left(\frac{1}{2}F(R)+3H\partial_{t}F'(R)\right)
 -3\dot{H}\, , \nn
 -3H^2-2\dot{H} &=&
 -\frac{1}{F'(R)}\left(\frac{1}{2}F(R)+2H\partial_tF'(R)
+\partial_{t}^2F'(R)\right)-\dot{H}\, ,
\label{3.2.1}
\eea
where we have neglected the contributions of any other kind of matter.
If we compare Eqs.~(\ref{3.2.1}) with the standard FRW equations
($3H^2=\kappa^2\rho$ and $-3H^2-2\dot{H}=\kappa^2p$),
we may identify both right sides of Eqs.~(\ref{3.2.1}) with the energy-density
and pressure of a perfect fluid,
in such a way that they are:
\bea
\rho &=& \frac{1}{\kappa^2}\left[ \frac{1}{F'(R)}
\left(\frac{1}{2}F(R)+3 H \partial_t F'(R)\right) -3\dot{H}\right] \nn
p &=& -\frac{1}{\kappa^2}\left[\frac{1}{F'(R)}\left(\frac{1}{2}F(R)
+2H\partial_t F'(R)+\partial_{t}^2F'(R)\right)+\dot{H}\right]\, .
\label{3.2.2}
\eea
Then, Eqs.~(\ref{3.2.1}) take the form of the usual FRW equations, where the
EoS parameter for this dark fluid is defined by:
\be
w=\frac{p}{\rho}=-\frac{\frac{1}{F'(R)}\left(\frac{1}{2}F(R)
+2H\partial_t F'(R)+\partial_{t}^2F'(R)\right)+\dot{H}}{ \frac{1}{F'(R)}
\left(\frac{1}{2}F(R)+3 H \partial_t F'(R)\right) -3\dot{H}}\, .
\label{3.2.3}
\ee
The corresponding EoS may be written as follows:
\be
p=-\rho-\frac{1}{\kappa^2}\left(4\dot{H}+\frac{1}{F'(R)}\partial_{t}^2F'(R)
 -\frac{H}{F'(R)}\partial_{t}F'(R)\right)\, .
\label{3.2.4}
\ee
The Ricci scalar is $R=6(2H^2+\dot{H})$, then $F(R)$ is a function of
the Hubble parameter $H$ and its derivative $\dot{H}$.
The form of the EoS is written as:
$p=-\rho+g(H,\dot{H}, \ddot{H}, \cdots)$, where
\be
g(H,\dot{H}, \ddot{H},
\cdots)=-\frac{1}{\kappa^2}\left(4\dot{H}+\partial^2_{t}(\ln F'(R))
+(\partial_t\ln F'(R))^2-H\partial_t\ln F'(R)\right)\, .
\label{3.2.5}
\ee
Then, by combining the FRW equations, it yields the following differential
equation:
\be
\dot{H}+\frac{\kappa^2}{2}g(H,\dot{H}, \ddot{H}, \cdots)=0\, .
\label{3.2.6}
\ee
Hence, for a given cosmological model, the function $g$ given in (\ref{3.2.5})
may be seen
as a function of cosmic time $t$, and then by the time-dependence of the Ricci
scalar,
the function $g$ is rewritten in terms of $R$. Finally, the function $F(R)$ is
recovered
by the expression (\ref{3.2.5}). In this sense, Eq.~(\ref{3.2.6}) combining
with the expression (\ref{3.2.5}) results in \cite{Saez-Gomez1}:
\be
\frac{dx(t)}{dt}+x(t)^2-H(t)x(t)=2\dot{H}(t)\, ,
\label{3.2.7}
\ee
where $x(t)=\frac{d\left(\ln F'(R(t)\right)}{dt}$.  In general, this equation is very difficult to be solved for a particular Hubble parameter, so that the reconstruction of $F(R)$ gravity is not so simple. Let us consider an example to illustrate this method, we assume a power-law evolution for the Hubble parameter,
\be
H(t)=\frac{\alpha}{t}\ .
\label{pw1}
\ee
Then, the solution of the equation \eqref{pw1} is given by,
\be
x(t)=\frac{\alpha}{2t}-\frac{-1+\sqrt{1-6\alpha+\alpha^2}\left(-1+\frac{2 C_1}{t^{\sqrt{1-6\alpha+\alpha^2}}+C_1}\right)}{2t}\ .
\label{pw2}
\ee
Recalling that $x(t)=\frac{d\left(\ln F'(R(t)\right)}{dt}$, the $F(R)$ action yields,
\be
F(R)=C_1\ R^{\frac{5+\alpha-\sqrt{1+\alpha(\alpha-6)}}{4}}\left(R^{\frac{\sqrt{1+\alpha(\alpha-6)}}{2}}+C_2\right)\ ,
\label{pw3}
\ee
where $\{C_1,C_2\}$ are integration constants. Hence, the $F(R)$ action has been reconstructed for the class of power-law solutions expressed in \eqref{pw1}. However, for a particular oscillating solution, as the ones studied in the section above, the equation (\ref{3.2.7}) turns out very complicated, and in general, one can not obtain an explicit expression for the action. Nevertheless, let us proceed in a different way, by reconstructing the Hubble parameter assuming a particular $x(t)$ in the equation \eqref{3.2.7}, 
\be
x(t)=\frac{d(\ln F'(R(t))}{dt}=H_1\cos\omega t\, .
\label{d6}
\ee
Then, by the equation (\ref{3.2.7}), the Hubble parameter obtained is given by,
\be
H(t)=C_1 \e^{-\frac{H1\sin\omega t}{2\omega}}+\e^{-\frac{H1\sin\omega t}{2\omega}}\int dt\ \e^{\frac{H1\sin\omega t}{2\omega}} \left(H_1^2+H1^2\cos\omega t-2H_1\omega\sin\omega t\right)\ .
\label{extra1}
\ee
And the action is obtained, 
\be
F(R)= \int dR\ \e^{-\frac{H_1}{\omega}\sin\left[ \omega t(R)\right]}\ .
\label{extra2}
\ee
Here the function $t(R)$ comes from the inversion of the definition of the Ricci scalar $R(t)=6(2H^2+\dot{H})$. Hence, this kind of cyclic evolution (\ref{extra1}) can be reproduced in the frame of $F(R)$ gravities. However, the complexity of the integral does not allow us to get an explicit expression for the $F(R)$ action. Note that for explicit values of the constant $H_1$, and of the frequency $\omega$, some information may be obtained for different times, since each term in the integral (\ref{extra1}) becomes important for different values of the time coordinate, i.e. at different epochs of the Universe evolution.

\section{The Ekpyrotic scenario in $F(R)$ gravity}

The inflationary epoch (early-time acceleration) just after the initial
singularity
that was the origin of the Universe, is able to solve some of intrinsic
problems of the Big Bang model.
These problems, known as the {\it flatness} problem, the question on the
homogeneity at large scales,
the absence of monopoles or the origin of the inhomogeneities that lead to the
formation of large scale
structure, can be solved quite well by the inflationary paradigm. However, a
decade ago a new proposal,
alternative to inflation, was suggested, the so-called Ekpyrotic/cyclic
Universe (see Ref.~\cite{ekpyrotic}).
This new scenario solves additionally the puzzles of the standard cosmological
model as well as it provides perhaps
a more complete picture of the Universe evolution. While in the inflationary
scenario, the initial conditions
are required to complete the cosmological picture, as the time begins at the
Big Bang singularity,
in the ekpyrotic Universe, this is not a requirement due to its cyclic nature.
The evolution presented
by this class of cyclic Universe contains in general four stages per cycle: a
first initial hot state
similar to the standard Big Bang model, then a phase of accelerated expansion,
after which a phase
where the Universe starts to contract occurs and finally ends in a Big
Bang/Crunch transition,
when the cycle starts again. The corresponding period where the main problems
enumerated above
are solved occurs during the contracting phase. In the usual ekpyrotic models,
a scalar field is
included to reproduce a cyclic Universe. However, it is clear that modified
gravity,
and precisely $F(R)$ gravity, can perfectly reproduce the ekpyrotic scenario.
In order to show
that the above models, reproduced by $F(R)$ gravity, can solve the initial
problems
by means of a cyclic Universe containing a contracting phase, let us consider a
FRW Universe described by,
\be
\frac{3}{\kappa^2}H^2=\frac{\rho_{m0}}{a^3}+\frac{\rho_{r0}}{a^4}
+\frac{\rho_{\sigma0}}{a^6}-\frac{k}{a^2}+\rho_{F(R)}\, .
\label{E1}
\ee
Here $\rho_{m0}$ is the energy-density for pressureless matter, $\rho_{r0}$
for radiation, $\rho_{\sigma0}$ for the anisotropies, $k$ is the spatial
curvature of
the Universe and $\rho_{F(R)}$ is the effective energy-density defined in the
previous section for the extra geometrical terms. Then, in order to get a
homogeneous and isotropic spatially
flat Universe in a contracting phase, the effective EoS parameter $w_{F(R)}>1$,
so that when
the scale factor tends to zero, the $F(R)$ terms in the equations dominate over
the rest,
and the result is the same as in the inflationary scenario.

Looking at the above examples, we have that for the model given in  (\ref{dd3}),
the effective EoS parameter for the perfect fluid defined with the extra terms
of $F(R)$ is given by,
\be
w_{F(R)}=\frac{p_{F(R)}}{\rho_{F(R)}}=-1+\frac{4\omega\left(-2\omega (\cos \omega t)+\omega (\sec \omega t)^2+H_0(\cos \omega t)(\sin \omega t)\right)}{3\left(H_0(\cos \omega t)^2+2\omega (\sin 2 \omega t)-(\tan \omega t)\right)^2}\, .
\label{E2}
\ee
Hence, for each epoch of the Universe, the effective EoS behaves in a different way, we can summarize the phases of each cycle as follows 
\be
w_{F(R)}=\left\{\begin{matrix}
 -1+\frac{8 \omega}{3H_0}+\frac{32 \omega^2}{3H_0^2}\ , & t\rightarrow \pi/4\omega \\
-\frac{2}{3}\ , &  t\rightarrow \pi/2\omega \\
-1-\frac{4\omega^2}{4H_0^2}\ , & t\rightarrow \pi/\omega 
\end{matrix}\right.
\label{ex2}
\ee
Hence, we can see that the $F(R)$ fluid dominates over the rest of components in \eqref{E2} when its EoS is larger than 1, which occurs for small times and a contracting universe, while for large times, the EoS behaves as dark energy, and even it crosses the phantom barrier when $w_{F(R)}<-1$. So that the ekpyrotic scenario
takes place, and the result is the observable Universe. Hence, we have shown
that cyclic Universe can be well reconstructed in frames of modified
gravity, and the initial
problems, as the {\it flatness} or {\it horizon} problems, can be solved.
However, generally
the cyclic Universes contain singularities of the type of Big Bang/Crunch,
where
the scale factor goes to zero,
and the energy densities for matter/radiation grow to infinity, as ,for
example,
occurs for the model (\ref{dd2}).
In the original ekpyrotic model, based on a scalar field, the divergences are
controlled by a function
introduced in the action as a strong coupling between the scalar field and the
other components.
In the kind of ekpyrotic scenario proposed here, a smooth transition along the
Big Bang/Crunch can
be reproduced by means of the reconstruction of models that avoid the
singularity (as it is realized
for several examples given above) or by introducing a coupling in the matter
action,
\be
S_M=\int dx^4 \sqrt{-g} \beta(R) \mathcal{L}_m\, .
\label{E3}
\ee
Hence, by an appropriate choice of the function $\beta(R)$, the matter
divergences can be avoided,
and the transition across the Big Bang/Crunch can be realized smoothly.

Therefore, the ekpyrotic scenario can be realized with no need to introduce an
additional field but only
in terms of modified gravity.

\section{Little rip cosmology in modified gravity}

In the previous sections, several oscillating Universes have been
reconstructed via $F(R)$ gravity.
Cyclic Universes seem to be well reproduced in these theories and the ekpyrotic
scenario can be realized.
Here we are interesting in the study of cosmological models which are able to
reproduce super-accelerated phase in
the context of modified gravity. In this case, the corresponding effective EoS
parameter (\ref{3.2.3}) crosses
the phantom barrier, that is $w_{F(R)}<-1$. It is well-known that cosmological
phantom models usually
contain the so-called big rip singularity \cite{rip}.
In the case of the big rip singularity, the phantom energy-density and the
scale factor diverges in a finite future. One of the most surprising
consequences of the big rip is the dissolution of bounded systems, as
the Solar System and atoms before the singularity (see
Ref.~\cite{rip,BreakCouplings}).

In this section,
we are interesting to reconstruct $F(R)$ gravity which is able to reproduce a
phantom behavior, being free of future singularity.
Nevertheless, such singularity free or
little rip cosmology (see Refs.~\cite{LittleRip,eli}) also leads to
dissolution of bound structures. Let us recall the first FRW equation in
$F(R)$,
\be
\frac{3}{\kappa^2}H^2(t)=\rho_{F(R)}(t)\, ,
\label{LR1}
\ee
where the energy-density is defined in (\ref{3.2.2}). By definition, a big rip
singularity occurs
in a finite time, usually denoted by $t_s$, when the scale factor $a(t)$ and
the energy-density
diverges. So in order to avoid a big
rip singularity,
$\rho_{F(R)}(t)$ must remain positive and finite for all $t$. Looking at the
expression for
$\rho_{F(R)}(t)$ in (\ref{3.2.2}), it seems complicated to obtain a general
condition on the form
of $F(R)$, but a natural condition (not sufficient) seems $F'(R)>0$ for all $R$
in order to avoid
divergences. On the other hand, in the absence of matter, the EoS parameter
(\ref{3.2.3}) can be rewritten as,
\be
w_{F(R)}=-1-\frac{2\dot{H}}{3H^2}\, .
\label{LR2}
\ee
Hence, for a super-accelerated expansion, the Hubble parameter is required to
be an increasing
function of time. Looking at some of the models reconstructed in
the previous
sections, we can see that the Hubble parameter (\ref{dd3}) is a
periodic function that
reproduces a cyclic Universe. From the effective EoS parameter (\ref{ex2}), it
is clear that this model
exhibits periods of super-accelerating expansion and it does not contain any
future singularity.
To find out if the model (\ref{dd3}) may lead to a little rip, one has to
determine the duration
of each cycle and the strength of the effective repulsive force reproduced by
the $F(R)$ terms
and compare with the binding forces of coupled systems (as for example the
Solar System). However,
to ensure that a little rip is reproduced, one has to study models of eternal
acceleration but free
of future singularities, which may have a stronger growth in time than those
models
containing big rip singularities. Let us use the reconstruction technique using
Eqs.~(\ref{PQR1}-\ref{XIV}). As
an example, we consider the function,
\be
U(\phi)=\e^{-\beta\e^{\alpha\phi}}\, ,
\label{LR3}
\ee
where $\alpha$ and $\beta$ are constants. Then, by the expression (\ref{XIV}),
the Hubble parameter and the scale factor yield,
\be
H(t)=h_0 \e^{\alpha t}+h_1\, , \quad \rightarrow \quad a(t)=a_0
\e^{4\beta\e^{\alpha t}+6\alpha t}\, .
\label{LR4}
\ee
where $h_0=4\alpha\beta$ and $h_1=6\alpha$. It is straightforward to see that
the function (\ref{LR4})
describes a Universe, where for small times $t\ll \alpha$, the Hubble parameter
can be approximated as
a constant, reproducing a de Sitter solution, as in the case of $\Lambda$CDM
model. For large times,
the Universe ends in an eternal phantom phase, where the EoS parameter
$w_{F(R)}<-1$, but without big rip
singularity. Nevertheless, a little rip (dissolution of bound structures)
might occur in a finite time, similarly to
the model presented in Ref.~\cite{LittleRip}, as it is pointed out below. The
functions $P(\phi)$ and
$Q(\phi)$ can be easily reconstructed by the expressions (\ref{eq:2.12}) and
(\ref{XIII}),
\be
P(\phi)=\e^{4\beta\e^{\alpha\phi}}\, , \quad Q(\phi)=-6\alpha^2(3+4\beta
\e^{\alpha\phi}) (3+8\beta\e^{\alpha\phi})\e^{4\beta\e^{\alpha\phi}}\, .
\label{LR5}
\ee
The $F(R)$ action that reproduces the solution (\ref{LR4}) can be
calculated
by inverting the expression of the Ricci scalar $R=6(2H^2+\dot{H})$ and by
using the expression (\ref{PQR4}), which yields,
\be
F(R)=\left[C_1 + C_2 \sqrt{4\frac{R}{R_0}+75}\right]
\e^{\sqrt{\frac{R}{12R_0}+\frac{25}{16}}} \, .
\label{LR6}
\ee
where $R_0=\alpha^2$, $C_1=-24\e^{-39/12}R_0$, and $C_2=2\sqrt{3}R_0$. Hence,
we have reconstructed
$F(R)$ model which is able to reproduce a super-accelerated (eternal) phase,
which does not lead to a future singularity. Note that
the action (\ref{LR6}) turns out to be the Einstein-Hilbert action plus some
corrections for small
values of the Ricci curvature $R$, where the exponential functions are expanded
in power series,
\be
F(R)\sim \kappa_1R+\kappa_2 \frac{R^2}{R_0}+\kappa_3 \frac{R^3}{R_0^2}
+ \cdots \, .
\label{LR7}
\ee
Here the couplings $\kappa_i$ are constants depending on $C_{1,2}$. Hence,
for small values of the Ricci scalar the action reduces to the action for
General Relativity
plus power-law curvature corrections, as $R^2$, which is known recipe to cure
the singularities and which has a viable behavior
(\cite{NonsingularFR}). Hence, the action (\ref{LR6}) represents a viable model
where GR can
be recovered while the curvature scalar corrections remain
small. It is important that such additional corrections become
relevant close to the little rip evolution.
In order to estimate the time for the little rip induced dissolution of bound
structures in a naive way, one might compare the energy-density of a bound
system as the Solar System with
the density $\rho_{F(R)}$. For the model (\ref{LR4}), such density can be
approximated for large times by
\be
\rho_{F(R)}=\rho_0\e^{2\alpha t}\, ,
\label{LR7a}
\ee
where $\rho_0$ is a constant that can be set by imposing that
the current value of the energy-density is
$\rho_{F(R)}(t_0)=\frac{3}{\kappa^2}H_0^2\sim 10^{-47}\, \mathrm{GeV}^4$, where
the age of the Universe
is taken to be $t_0\sim 13.73\, \mathrm{Gyrs}$, according to
Ref.~\cite{Spergel:2006hy}. One can set the time
of the little rip dissolution occurrence when the
gravitational coupling of the Sun-Earth system is broken due to the
cosmological expansion. By assuming a mean density of the Sun-Earth system
given
by $\rho_{\odot-\oplus}=0.594\times 10^{-3}\, \mathrm{kg/m}^3 \sim 10^{-21}\,
\mathrm{GeV}^4$,
the time for the little rip dissolution of bound structures is,
\be
t_\mathrm{LR}=13.73\, \mathrm{Gyrs} + \frac{29.93}{\alpha}\, .
\label{LR7b}
\ee
Hence, depending on the parameter $\alpha$, the appearance of the little rip
may last shorter or longer. For example, when $\alpha=10^{-1}\,
\mathrm{Gyrs}^{-1}$, the little rip occurs
at the Universe age of $t_\mathrm{LR}=313.03\, \mathrm{Gyrs}$, while for
$\alpha\geq 1\, \mathrm{Gyrs}^{-1}$,
the time for the decoupling will be much shorter.

A lot of models can be reconstructed in frames of $F(R)$ gravity which are able
to reproduce a
super-accelerating phase free of singularities. Let us consider, for example, a
Hubble parameter
that reproduces a phantom Universe without big rip \cite{Nojiri:2009kx},
\be
H^2=\frac{H_0^2}{4}(t-t_0)^2=H_0 N+H_1\, .
\label{LR8}
\ee
Here we have introduced the number of e-foldings $N=\ln\frac{a}{a_0}$ instead
of the cosmological time $t$.
For the Universe described by the function (\ref{LR8}), the effective EoS
parameter is less than
$-1$, so it also describes a phantom evolution but free of future
singularities. By using
the reconstruction technique from Ref.~\cite{Nojiri:2009kx}, the action that
reproduces
the above solution is given by,
\be
F(R)=K\left(-2,-\frac{1}{2};\frac{R-3H_0}{12H_0}\right)\, ,
\label{LR9}
\ee
where $K(a,b,x)$ is the Kummers serie. In this model, the energy-density is
given by,
\be
\rho_{F(R)}=\rho_0 (t-t_0)^2\, ,
\label{LR9a}
\ee
where $t_0=0$ as the origin of the Universe evolution. Then, by adjusting
the value for $\rho_0$ with the present value of
the energy-density as in the above model,
the little rip occurs when the Universe age is
$t=137.3\times 10^{12}\, \mathrm{Gyrs}$, which clearly
grows much slower than for above model.

Hence, we have shown here that $F(R)$ gravity is able to reproduce successfully
a phantom scenario
free of future singularity, where the little rip occurs at the time which
depends completely on the expansion growth rate of the model.

We are now interested to explore the corresponding picture in the Einstein
frame for those solutions
describing a little rip in $F(R)$ gravity. In order to reconstruct the action
in the Einstein frame,
one has to use a conformal transformation that removes the strong coupling in
the action (\ref{PQR1}),
\be
g_{E\mu\nu}= \Omega^2g_{\mu\nu}\, , \quad \text{where} \quad \Omega^2=
P(\phi)\, ,
\label{CT1}
\ee
where the subscript $_\mathrm{E}$ stands for Einstein frame. A
quintessence-like action results in the Einstein frame
\be
\label{CT1b}
S_\mathrm{E}=\int d^4x \sqrt{-g_\mathrm{E}} \left[R_\mathrm{E} -\frac{1}{2}
\omega(\phi) \partial_{\mu}\phi
\partial^{\mu}\phi -U(\phi) \right]\, ,
\ee
where
\be
\omega(\phi)=\frac{12}{P(\phi)}\left(
\frac{d\sqrt{P(\phi)}}{d\phi}\right)^2\, ,
\quad U(\phi)=\frac{Q(\phi)}{P^2(\phi)}\, ,
\label{CT2}
\ee
while the cosmological solutions in the Jordan frame are transformed through
out the conformal transformation as,
\be
a_\mathrm{E}(t_\mathrm{E})=P(\phi(t))^{1/2}a(t)\, \quad \text{where} \quad
dt_\mathrm{E}=P(\phi(t))^{1/2}dt\, .
\label{CT3}
\ee
As an example, one can consider the action (\ref{LR5}) and (\ref{LR6}), which
was found to
reproduce the solution (\ref{LR4}). By applying the conformal transformation
(\ref{CT1}),
the action in the Einstein frame (\ref{CT2}) is described by the kinetic term
and the scalar potential,
\be
\omega(\phi)=4\alpha^2\beta^2\e^{2\alpha\phi}\, , \quad \mbox{and}
\quad U(\phi)=-6\alpha^2(3+4\beta
\e^{\alpha\phi})(3+8\beta\e^{\alpha\phi})\e^{-4\beta\e^{\alpha\phi}}\, .
\label{CT3B}
\ee
The solution is transformed as,
\be
a_\mathrm{E}(t)=a_0 \e^{6(\beta\e^{\alpha t}+\alpha t)}\, , \quad
t_\mathrm{E}=\int_{-\infty}^{2\beta\e^{\alpha t}}\frac{\e^{z}}{z}\, .
\label{CT4}
\ee
Here the second expression is given by an exponential integral. In order to
calculate an exact
expression for $a_\mathrm{E}(t_\mathrm{E})$, let us approximate the exponential
$\e^{\alpha t}\sim 1+\alpha t + O(t^2)$
for small times, so that the solution can be expressed as,
\be
a_\mathrm{E}(t_\mathrm{E})=a_{E0} t_\mathrm{E}^6 \e^{6\beta\gamma
t_\mathrm{E}}\, ,
\quad H_\mathrm{E}(t_\mathrm{E})= 6\beta\gamma +\frac{6}{t_\mathrm{E}}\, ,
\label{CT5}
\ee
where $\gamma=\alpha \e^{-2\beta}$. Then, it is straightforward to see that the
Universe described
in the Einstein frame not only does not describe a super-accelerating expansion
evolution but it
also contains a singularity, a type of initial or Big Bang singularity. On the
same time the solution in
the Jordan frame was free of singularities. The singularity of the Einstein
frame universe is probably the manifestation of little rip (dissolution of
bound structures).

Let us consider now the inverse reconstruction by studying first the little rip
scenario with
quintessence/phantom fields described by the action (\ref{ma7}), and then
reconstruct the $F(R)$
action by means of a conformal transformation. It is well-known that any
cosmological solution can
be reconstructed with minimally coupled quintessence/phantom fields scalar
fields
(see Ref.~\cite{Elizalde:2008yf}). As an example, let us consider the kinetic
term and the scalar potential
\be
\omega(\phi)=-\frac{2}{\kappa^2}h_0\alpha\e^{\alpha \phi}\, , \quad
V(\phi)=\frac{1}{\kappa^2}\left(3h_0^2\e^{2\alpha
\phi}+(6h_0h_1+h_0\alpha)\e^{\alpha \phi}+3h_1^2\right)\, .
\label{LR10}
\ee
Using the FRW equations and Eqs.(\ref{ma10}), the solution for the Hubble
parameter and the scalar field are
\be
H_\mathrm{E}(t)=h_0 \e^{\alpha t_\mathrm{E}}+h_1\, , \quad
\phi(t_\mathrm{E})=t_\mathrm{E}\, ,
\label{LR11}
\ee
That coincides with the solution found in (\ref{LR4}) reproducing the same
little rip evolution
as in $F(R)$ gravity. In this case the equation of state parameter is,
\be
w_{\phi}=\frac{\frac{1}{2}\omega(\phi)\dot{\phi}^2-V(\phi)}{\frac{1}{2}\omega(\phi)\dot{\phi}^2+V(\phi)}
=-1-\frac{2h_0\alpha}{3(h_0\e^{\frac{\alpha}{2} t_\mathrm{E}}+h_1\e^{-\alpha
t_\mathrm{E}})^2}\, .
\label{LR12}
\ee
Clearly this EoS parameter describes a scalar field which has a phantom
behavior, where its EoS parameter
$w_{\phi}<-1$ and only tends to $-1$ for $t_\mathrm{E}\rightarrow\infty$.
However, as a novelty compared with
the usual phantom scalar fields (Ref.~\cite{Elizalde:2008yf}), here there is no
finite time future singularity,
but little rip evolution occurs. For this example, the time for the
dissolution of
the Sun-Earth system is given by (\ref{LR7b}).

Let us now attempt to reconstruct the corresponding $F(R)$ by applying
a conformal transformation to the action (\ref{ma7}). It is
well-known that $F(R)$ gravity can be described
by a non-propagating scalar field (\ref{PQR1}), so that the conformal
transformation must remove the scalar
kinetic term in the action (\ref{ma7}):
\be
g_{\mu\nu E}=\Omega^2 g_{\mu\nu}\, , \quad \mbox{where} \quad
\Omega^{2}=\exp\left[\pm\sqrt{\frac{2}{3}} \kappa \int
d\phi\sqrt{\omega(\phi)}\right]\, ,
\label{LR13}
\ee
The corresponding Jordan frame action yields,
\be
S=\int d^4x \sqrt{-g} \left[ \frac{\e^{\left[\pm \sqrt{\frac{2}{3}}\kappa
\int d\phi\sqrt{\omega(\phi)}\right]}}{2\kappa^2}R-\e^{\left[\pm
2\sqrt{\frac{2}{3}}\kappa
\int d\phi\sqrt{\omega(\phi)}\right]} V(\phi) \right] \, ,
\label{LR14}
\ee
This action is basically the action given in (\ref{PQR1}) in vacuum, which is
equivalent to $F(R)$
gravity. However, for the case of a phantom scalar that reproduces the little
rip, it is
straightforward to see that the corresponding action (\ref{LR14}) turns out to
be complex, as the conformal
transformation (\ref{LR13}) is also complex due to fact that kinetic term in
(\ref{LR12}) is negative. This is
the known result for the case of a phantom scalar cosmology containing future
singularity (see
Ref.~\cite{Briscese:2006xu}). As in the case of phantom
singular Universes, for the case of little rip, a way to reconstruct a
consistent action in the Jordan
frame can be achieved by adding an extra phantom fluid (see
Ref.~\cite{Elizalde:2009gx}). However,
in general a little rip Universe described in the Einstein frame owns an
inconsistent action in the Jordan frame.
Thus, it is explicitly demonstrated that little rip evolution may be
reconstructed in frames of modified gravity.

\section{Discussion}

In summary, we developed the reconstructed method for $F(R)$ theory which
admits cyclic and ekpyrotic universe solution. As a rule, the
corresponding reconstructed $F(R)$ action is given implicitly. Nevertheless,
its expansion around General Relativity action with curvature
corrections is always possible as is shown explicitly. The comparison with
scalar-tensor theory which
also leads to cyclic evolution is made. Our results indicate that
geometrical
actions usually considered as inflation source may lead to more
complicated cyclic universe.

We also presented the reconstruction of $F(R)$ theory leading to little
rip universe. Little rip dark energy represents non-singular phantom
cosmology. It was proposed in Ref.~\cite{LittleRip} as viable alternative
to $\Lambda$CDM model because it is consistent with observational data.
Being effectively non-singular one, little rip cosmology shows the
dissolution of bound structures similarly to Big Rip cosmology.
It is demonstrated that same effect occurs for $F(R)$ gravity admitting
little rip solution: the corresponding time left for such dissolution is
estimated. It is interesting that transforming little rip $F(R)$ gravity
to scalar-tensor form (the Einstein frame), the little rip universe is
transformed to qualitatively different (Big Bang type) singular universe.
The appearance of initial singularity in Einstein frame indicates the
manifestation of dissolution time in Einstein frame.
Finally, it is shown that other types of non-singular super-accelerating
universe may be
also reconstructed in $F(R)$ gravity. Note also that the kind of solutions studied here can be also extended to Ho\v{r}ava-Lifshitz gravity (see Ref.~\cite{LopezRevelles:2012cg}).

Note that modified gravity theories under discussion for cyclic universe 
are relevant mainly for
early universe and/or may be expanded as General Relativity action plus
corrections. Hence, such theories easily pass the local tests. From another
side, it remains to understand how natural and realistic the
cyclic/ekpyrotic/little rip universes are. The natural recipe for this 
purpose
is the study of perturbations. However, such investigation should be done for
different epochs showing subsequent transition from the earlier epoch to the
following oneand also with account of corresponding matter/radiation. This 
is quite
complicated technical problem which lies outside of the scopes of current work.

\section*{Acknowledgments}

We are grateful to J. Khoury and R. Scherrer for useful remarks.
S.N. was supported in part by Global COE Program of Nagoya University (G07)
provided by the
Ministry of Education, Culture, Sports, Science \& Technology ; the
JSPS Grant-in-Aid for Scientific Research (S) \# 22224003 and (C) \# 23540296.
S.D.O. was supported by MICINN (Spain) projects FIS2006-02842 and
FIS2010-15640,
by CPAN Consolider Ingenio Project and
AGAUR 2009SGR-994 and by AEI, Golm. DSG acknowledges a FPI fellowship from
MICINN (Spain), project FIS2006-02842.
S.D.O. is grateful to H.Nicolai for warm hostipality at AEI, Golm where this
work was completed.

\end{document}